# Analyzing conformational changes in single FRET-labeled A$_1$ parts of archaeal A$_1$A$_O$-ATP synthase

Hendrik Sielaff[a,b,#], Dhirendra Singh[b,#], Gerhard Grüber[b,*] Michael Börsch[a,c,*]

[a] Jena University Hospital, Single-Molecule Microscopy Group, Nonnenplan 2 - 4, 07743 Jena, Germany;
[b] Nanyang Technological University, School of Biological Sciences, 60 Nanyang Drive, Singapore 637551, Republic of Singapore
[c] Center of Medical Optics and Photonics (CeMOP), Jena, Germany

## ABSTRACT

ATP synthases utilize a proton motive force to synthesize ATP. In reverse, these membrane-embedded enzymes can also hydrolyze ATP to pump protons over the membrane. To prevent wasteful of ATP hydrolysis, distinct control mechanisms exist for ATP synthases in bacteria, archaea, chloroplasts and mitochondria. Single-molecule Förster resonance energy transfer demonstrated that the C-terminus of rotary subunit ε in *Escherichia coli* changes its conformation to block ATP hydrolysis. Previously, we investigate the related conformational changes of subunit F of the A$_1$A$_O$-ATP synthase from the archaeon *Methanosarcina mazei* Gö1. Here, we analyzed the lifetimes of fluorescence donor and acceptor dyes to distinguish between signals and potential artefacts.

## 1. INTRODUCTION

Single-molecule Förster resonance energy transfer (smFRET) is a powerful tool to measure inter-domain motions in proteins, for example the rotary motion of subunits in F$_1$F$_O$-ATP synthase[1-11]. Different sites of a protein can be labeled specifically with a donor and an acceptor fluorophore. Exciting the donor fluorophore may result in energy transfer to the nearby acceptor dye. According to T. Förster[12-14] the energy transfer efficiency between a donor and an acceptor fluorophore depends on several factors, e.g. the distance, the transition dipole moment orientations, the fluorescence quantum yield, the fluorescence lifetime of the donor or quenching processes of the fluorophores by the local environment. The donor excited state lifetime of a FRET-labeled protein depends on two competing radiative processes beside non-radiative decay pathways, i.e. the fluorescence decay and the energy transfer in the presence of an acceptor dye. Therefore, the fluorescence lifetime of the donor is expected to decrease in the presence of an acceptor due to FRET. On the other hand, the acceptor fluorescence lifetime remains unchanged when compared to the lifetime of the acceptor dye *via* direct excitation. Measuring the fluorescence lifetime of the donor fluorophore can serve as an independent control for FRET in cuvette-based experiments or in confocal microscopy of single molecules.

Here, we investigated the cytoplasmic domain of A-ATP synthase from the archeon *Methanosarcina mazei* Gö1 (*M. mazei* Gö1) and analyzed the fluorescence lifetime changes related to smFRET. The A-ATP synthase consists of a membrane-embedded, proton conducting A$_O$ portion and a soluble A$_1$ portion that uses the energy generated by a transmembrane electrochemical potential of H$^+$/Na$^+$ to synthesize ATP from ADP and Pi[15,16]. Our soluble protein complex comprised subunits A$_3$B$_3$DF of the A$_1$ portion (Fig. 1A) that can only function in the direction of ATP hydrolysis and is incapable of synthesizing ATP. Subunits D and F form a central stalk that connects the A$_1$ portion with the A$_O$ portion in the complete enzyme[17,18]. The tip of subunit D forms a coiled-coil fold that inserts into the hexagonal A$_3$B$_3$-headpiece[17,19]. Together with subunit F it serves as a rotating nanomotor that moves in 120° steps[20] driven by ATP hydrolysis in the three nucleotide-binding sites formed by the interface of three pairs of subunits A and B.

___

*email: michael.boersch@med.uni-jena.de; ggruber@ntu.edu.sg; [#]Authors contributed equally

During ATP hydrolysis the flexible C-terminal domain (CTD) of subunit F is believed to move up and down[21,22]. Recently, we measured the conformational change of the CTD in relation to subunit D by smFRET[23], whereby a cysteine at the bottom of subunit D ($D_{A71C}$) was labeled with the FRET donor fluorophore Atto488, and a second cysteine in the CTD of subunit F ($F_{L87C}$) was labeled with the FRET acceptor fluorophore Atto647N. We observed changes of the FRET efficiency in single $A_3B_3DF$ complexes in the presence of Mg-ATP. Here, we investigated the donor and acceptor lifetimes in single $A_3B_3DF$ molecules to reveal the conformational changes as indicated by the intensity-based smFRET analysis.

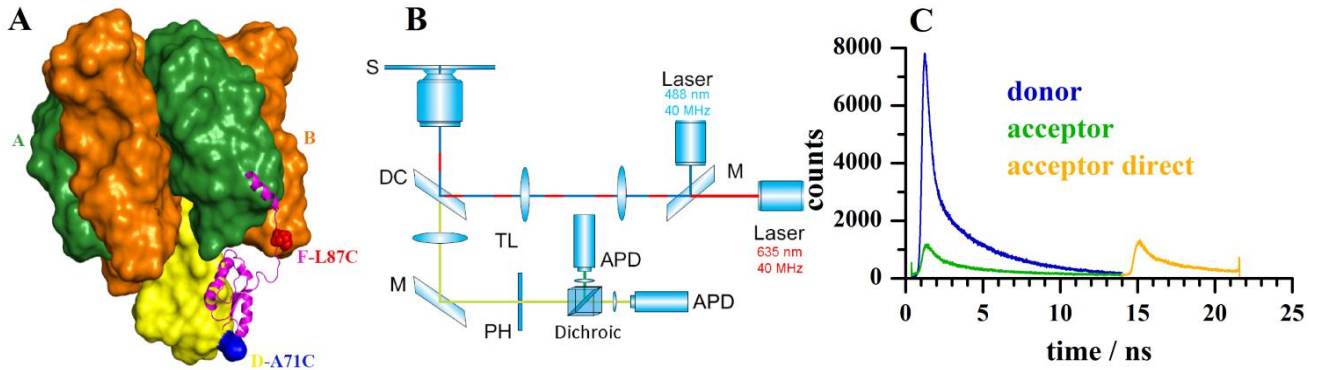

**Figure 1**: **A)** Model of the $A_3B_3DF$ complex of $A_1A_O$-ATP synthase from archaea *M. mazei* Gö1. The FRET donor fluorophore position $D_{A71C}$ (blue sphere on subunit D in yellow) and the FRET acceptor fluorophore position $F_{L87C}$ (red sphere on subunit F in magenta) are highlighted. Subunits A and B are orange and green, respectively. **B)** Confocal microscope setup for smFRET using pulsed duty-cycle optimized alternating laser excitation with 488 nm and 635 nm. **C)** Time-resolved fluorescence decays for FRET donor Atto488 (blue decay) and acceptor Atto647N (green decay) upon excitation with 488 nm as well as acceptor decay following direct excitation with 635 nm (orange decay, pulse delayed by 14 ns).

## 2. EXPERIMENTAL PROCEDURES

### 2.1 Preparation of the FRET-labeled $A_3B_3DF$ complex

Cloning, expression, purification and labeling of cysteines in the *M. mazei* Gö1 $A_3B_3D_{A71C}$ complex and subunit $F_{L87C}$ was performed as described[23]. Labeled $A_3B_3D$ complexes and subunit F were recombined by stoichiometric incubation at 1 µM for 30 min in buffer A (50 mM Tris, pH 7.5, 100 mM NaCl, 4 mM $MgCl_2$) immediately before the smFRET experiments to yield the FRET-labeled $A_3B_3DF$ complex at nanomolar concentrations.

### 2.2 Custom-designed confocal microscope for smFRET

Time-resolved smFRET measurements were performed on a custom-designed confocal microscope (Fig. 1B, Olympus IX 71) equipped with a 60x water immersion objective (UPlanSApo, N.A. 1.2, Olympus). A ps-pulsed laser (PicoTA 490, Picoquant, Germany) excited the FRET donor Atto488 (ATTO-TEC, Germany) with 150 µW at 488 nm using a repetition rate of 40 MHz. A second pulsed laser diode (LDH 635B, Picoquant, Germany) was used for pulsed duty-cycle optimized alternating excitation[24-31] of the FRET acceptor Atto647N (ATTO-TEC, Germany) with 30 µW at 635 nm. The pulse was delayed by 14 ns with respect to the preceding 488 nm pulse (Fig. 1C). A dual-band dichroic beam splitter (HC dual line 488/633-638, AHF, Germany) blocked scattered light. Fluorescence passing a 150 µm pinhole was separated into two spectral channels by a dichroic beam splitter (zt 640 RDC, AHF, Germany). FRET donor fluorescence was detected between 500 and 570 nm using a band pass filter (ET 535/70M, AHF, Germany). FRET acceptor fluorescence was detected for wavelengths $\lambda$ > 647 nm by a combination of two long pass filters (Edge Basic LP 635, Razor Edge LP 647 RU, AHF, Germany). Two single photon-counting avalanche photodiodes (SPCM-AQRH-14, Perkin-Elmer) were used for simultaneous spectrally-resolved FLIM[32]. Two out of four synchronized TCSPC cards (SPC 154, Becker&Hickl, Germany) recorded the photons[8,9,28]. The Burst_Analyzer software (Becker&Hickl, Germany) was used to visualize time-resolved photon traces of single molecules, for selecting photon bursts based on intensity thresholds, and for calculating histograms of proximity factors and other photophysical properties.

# 3. RESULTS

## 3.1 Protein labelling

The CTD of subunit F is enhancing ATP hydrolysis activity in the *M. mazei* Gö1 $A_3B_3DF$ complex. It has been shown by NMR spectroscopy that the CTD of soluble subunit F is flexible and can exist in a retracted or extended form[22]. To investigate the conditions for a movement of subunit F we designed a smFRET experiment with a mutant $A_3B_3D$ complex that carried two cysteines for fluorescence labeling[23]. The cysteine in subunit D ($D_{A71C}$) was labeled with the FRET donor Atto488-maleimide and the cysteine in the CTD of subunit F ($F_{L87C}$) was labeled with the FRET acceptor Atto647N-maleimide. Labeling efficiencies of 72% and 100% were achieved for the FRET donor and acceptor on subunits D and F, respectively.

## 3.2 Measuring smFRET with $A_3B_3DF$ complexes in solution

Cysteine positions were chosen for a distance of about 4 nm assuming the retracted conformation of subunit F in the $A_3B_3DF$ complex. This distance matches the Förster radius $R_0$ (5.1 nm according to the supplier Atto-Tec GmbH, for a 50% energy transfer efficiency) of the FRET fluorophore pair Atto488-Atto647N including the short linkers. Accordingly this is the optimal distance for measuring small conformational changes of the CTD of subunit F by smFRET with the highest sensitivity.

FRET-labeled $A_3B_3DF$ complexes were obtained by stoichiometric mixing of Atto488-labeled $A_3B_3D$ with Atto647N-labeled subunit F at a final concentration of 1 µM each. A 50 µl droplet of a diluted nM protein solution was placed on a cover glass, and the alternating lasers were focused about 100 µm deep into the protein solution. When labeled proteins diffused through the detection volume photon bursts were generated. Photons were registered for 1000 s, and time traces were binned in 2 ms intervals. FRET-labeled $A_3B_3DF$ complexes were identified in the time traces by an automated search. We applied intensity thresholds for a photon burst, i.e. minimum intensities for the FRET donor following pulsed excitation with 488 nm as well as for the FRET acceptor directly excited by the 635 nm pulses. In addition, the minimal photon burst length was set to 5 ms, i.e. ten times longer than the average diffusion time of a single fluorescent impurity in buffer.

Fig. 2 shows different FRET efficiencies found in photon bursts from individual FRET-labeled $A_3B_3DF$ complexes measured in the presence 1 mM Mg-ATP. The different relative intensities of FRET donor, $I_D$, (Fig. 2, blue traces) and FRET acceptor, $I_A$, (Fig. 2, green traces) upon excitation with 488 nm were used to calculate the simplified equivalent of a FRET efficiency, i.e. the proximity factor, P, for each time bin within a photon burst (Fig. 2, red traces), using equation (1):

$$P = I_A / (I_D + I_A) \tag{1}$$

Both intensities were corrected for a background signal but not for spectral detection efficiencies nor the slightly different fluorescence quantum yields of Atto488 ($\Phi_{fl} \sim 0.8$) and Atto647N ($\Phi_{fl} \sim 0.65$). For each of the four photon bursts, the signal of the FRET acceptor following direct excitation with 635 nm was shown as the orange trace in Fig. 2A-D. In addition, the fluorescence lifetimes of the FRET donor together with monoexponential decay fit were plotted (Fig. 2E-H). While the proximity factor of the two photon bursts in Fig. 2A and 2B were very similar and around P = 0.5 ("*high FRET*"), the corresponding FRET donor fluorescence lifetimes were very different, i.e. either a long lifetime $\tau_{DA}(h)_{long} = 3.1$ ns (Fig. 2E) or a short lifetime $\tau_{DA}(h)_{short} = 1.1$ ns (Fig. 2F). Similarly, the two photon bursts in Fig. 2C and 2D exhibited a similar mean low proximity factor P ~ 0.2 ("*low FRET*"), but the corresponding FRET donor fluorescence lifetimes were either long ($\tau_{DA}(l)_{long} = 3.3$ ns, Fig. 2G) or short ($\tau_{DA}(l)_{short} = 1.9$ ns, Fig. 2H).

We also found photon bursts that yielded fluorescence solely from the FRET donor fluorophore, i.e. from protein complexes without a labeled subunit F, or with a bleached acceptor. Fig. 3A-C shows three typical "*donor only*" bursts (blue traces), where no acceptor signals (green and orange traces) were detected. Accordingly, the respective proximity factors were P = 0. For these "*donor only*" bursts we found distinct lifetimes upon monoexponential fitting: a long ($\tau_D(long)$ = 3.9 ns, Fig. 3D), a medium ($\tau_D(medium) = 2.4$ ns, Fig. 3E), and a short ($\tau_D(short) = 0.9$ ns, Fig. 3F) lifetime. These varying lifetimes corroborated the different $\tau_{DA}$ lifetimes found for FRET-labeled $A_3B_3DF$ complexes with the same proximity factor. In contrast, for $A_3B_3DF$ complexes without a FRET donor on subunit D (due to limited labeling efficiency or photobleaching) we observed photon burst that were characterized by fluorescence from the directly excited FRET acceptor only.

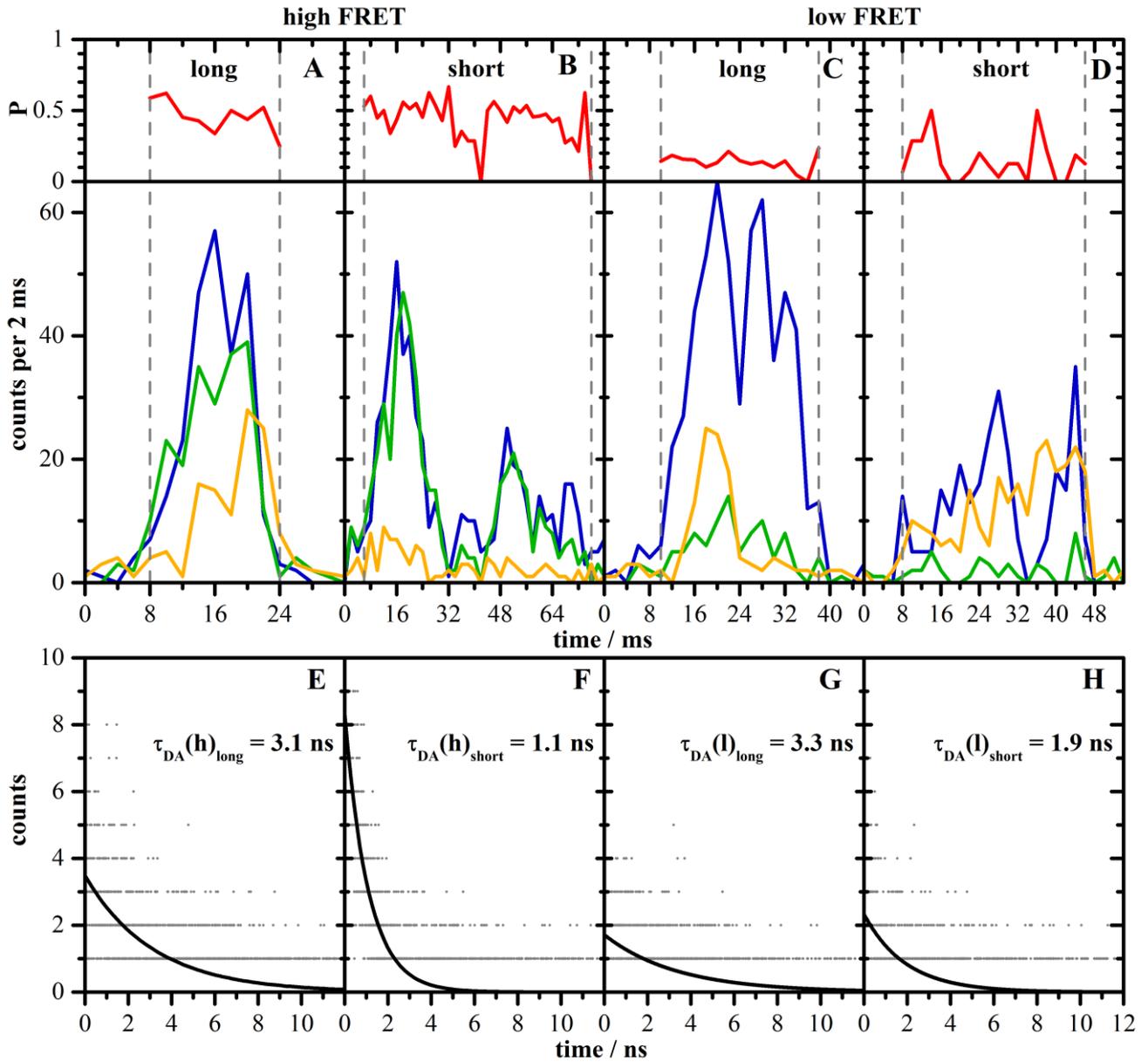

**Figure 2: A-D)** Single photon bursts of double-labeled $A_3B_3DF$ complexes in the presence of 1 mM Mg-ATP with the respective proximity factor P (top panels, red traces). Proximity factors were calculated for each 2 ms time bin from FRET donor intensities (blue traces) and FRET acceptor intensities (green traces) and classified as "*high FRET*" (**A, B**) or "*low FRET*" (**C, D**). FRET acceptor fluorophore intensities following direct excitation with 635 nm are shown as orange traces. **E-H)** Corresponding FRET donor lifetimes $\tau_{DA}$ were fitted as monoexponential decays.

For comparison we also measured photon bursts from individual FRET-labeled $A_3B_3DF$ complexes in the absence of Mg-ATP (buffer only) that exhibited a distribution of proximity factors with a peak between P = 0.4–0.75 ("*high FRET*"). We obtained the same result when the buffer contained 1 mM Mg-ADP or 1 mM Mg-AMPPNP, a non-hydrolysable Mg-ATP derivate. In the presence of 1 mM Mg-ATP we observed an additional second peak in the proximity factor distribution with P = 0.1–0.25 ("*low FRET*"), indicating an increase of the distance between donor and acceptor[23] (see below).

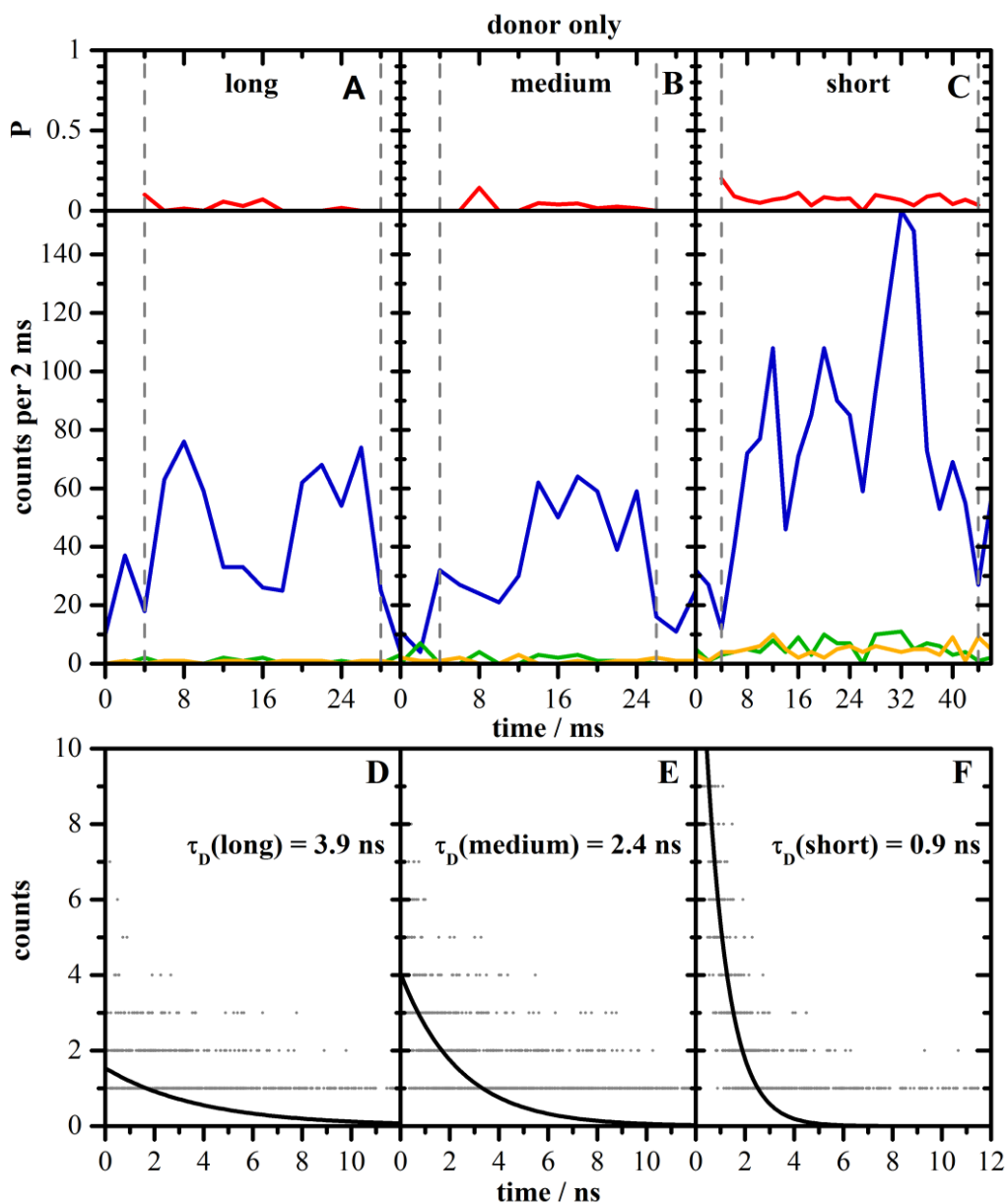

**Figure 3: A-C)** Single photon bursts of donor-labeled $A_3B_3DF$ complexes without acceptor dye in the presence of 1 mM Mg-ATP with the respective proximity factor P (top panels, red traces). Proximity factors were calculated for each 2 ms time bin from FRET donor intensities (blue traces) and FRET acceptor intensities (green traces). FRET acceptor fluorophore intensities following direct excitation with 635 nm were shown as orange traces. **D-F)** Corresponding "*donor only*" lifetimes, $\tau_D$, were fitted as monoexponential decays.

## 3.3 Single-molecule fluorescence lifetime measurements

We analyzed the FRET donor lifetimes of a sub-set of photon bursts for complexes with a FRET acceptor fluorophore attached (and in the presence of 1 mM Mg-ATP) that exhibited a minimal burst length of 8 ms and a maximum standard deviation for the proximity factor within the burst of 0.12. Thereby, more than 80% of all photon bursts were omitted from the combined FRET donor lifetime analysis. The remaining bursts were sorted according to the proximity factor into two categories, i.e. for P = 0.1-0.25 ("*low FRET*") or for P = 0.4–0.75 ("*high FRET*"). As shown in Fig. 4 adding all photons of bursts of one category resulted in lifetime histograms that had to be fitted by a double-exponential decay, i.e. for "*high*

*FRET*",τ(h), as well as for "*low FRET*",τ(l). The average short and long lifetimes were τ(h)$_{short}$ = 0.6 ± 0.02 ns and τ(h)$_{long}$ = 3.9 ± 0.09 ns for "*high FRET*" photon bursts (Fig. 4A), or τ(l)$_{short}$ = 0.5 ± 0.02 ns and τ(l)$_{long}$ = 3.9 ± 0.07 ns for "*low FRET*" photon bursts (Fig. 4B), respectively.

The lifetime histograms were compared to those lifetimes of complexes with "*donor only*" fluorescence (Fig. 4C). These photon burst were selected from molecules that showed a strong FRET donor fluorophore intensity with at least 50 photons on average, but no acceptor signal, i.e. less than 6 photons on average in the acceptor channel. We found distinct short or long fluorescence lifetimes τ(D) for these "*donor only*" bursts, which we used to determine τ(D)$_{short}$ = 0.7 ± 0.02 ns and τ(D)$_{long}$ = 4.2 ± 0.07 ns, respectively. The long lifetime corresponded to the published lifetime of the fluorophore Atto488 in solution (τ = 4.1 ns, according to the supplier Atto-Tec GmbH). In addition, we calculated the fluorescence lifetime of the FRET acceptor fluorophore by direct excitation with 635 nm (Fig. 3D), i.e. from photon bursts of both "*low FRET*" and "*high FRET*" proximity factor. The monoexponential decay fit yielded τ(A) = 3.6 ± 0.06 ns, which was the similar to the published fluorescence lifetime of the fluorophore Atto647N in solution (τ = 3.5 ns, Atto-Tec GmbH).

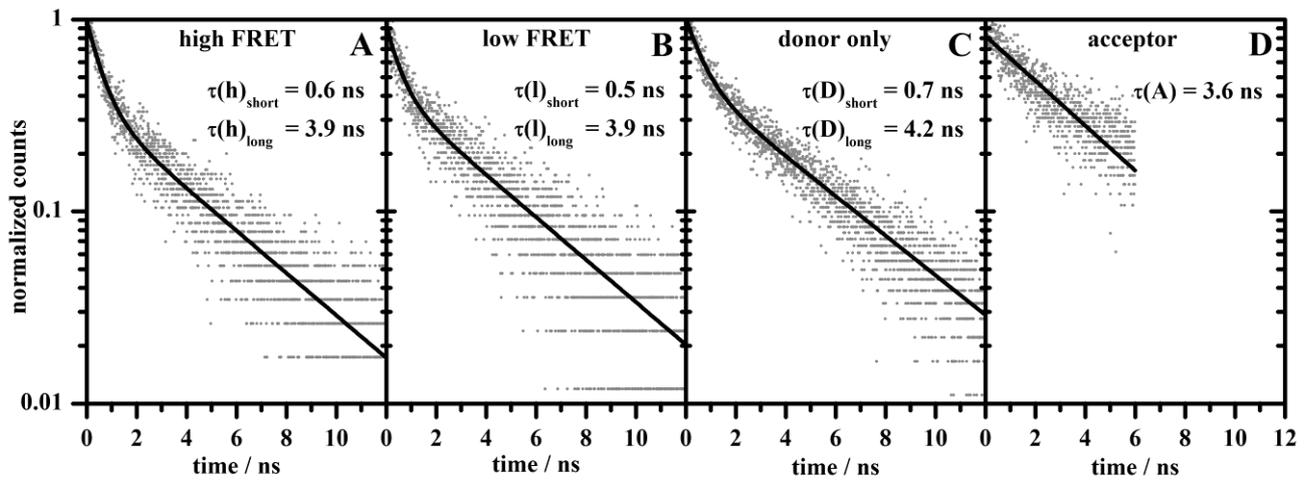

**Figure 4:** Combined normalized fluorescence lifetime decays from single photon bursts of the FRET-labeled A$_3$B$_3$DF complex and double-exponential decay fittings. Combined decays resulted from **A)** photons from 42 bursts that were classified as "*high FRET*", **B)** photons from 58 bursts classified as "*low FRET*", **C)** photons from 40 burst classified as "*donor only*", and **D)** photons from 49 bursts upon direct excitation of the acceptor fluorophore.

### 3.4 2D-distributions of mean intensities *versus* proximity factor

Both single burst-integrated lifetimes as well as the combined fluorescence lifetime analysis showed that the local protein environment for Atto488 at subunit D caused a partial fluorescence quenching. However, to ensure that the observed additional "*low FRET*" population of A$_3$B$_3$DF complexes in the presence of Mg-ATP is due to an induced conformational change of the CTD of subunit F, we also evaluated the possibility of partial FRET acceptor quenching. We assumed that a possible Mg-ATP quenching of the FRET acceptor could result in an apparent low FRET efficiency according to equation (1).

To validate or disprove this hypothesis we plotted the mean intensity of the FRET donor and FRET acceptor versus the mean proximity factor of the photon burst (Fig. 5). In general, the mean intensity of the FRET donor should decrease linearly with increasing P according to equation (1) for an undisturbed FRET relation. *Vice versa*, the mean intensity of the FRET acceptor should increase linearly with increasing P. In the presence of 1 mM Mg-ATP (Fig. 5A,E), or 1 mM Mg-ADP (Fig. 5C,G), or 1 mM AMPPNP (Fig. 5D,I), as well as in the absence of nucleotides (Fig. 5B,F), no obvious deviation from these linear relations were observed. For all biochemical conditions the mean FRET donor intensities for the lower limit of P = 0 were identical and approached 13 to 17 counts per ms, and the mean FRET acceptor intensities for the upper limit of P = 1 were almost identical and approached 12 to 17 counts per ms as well.

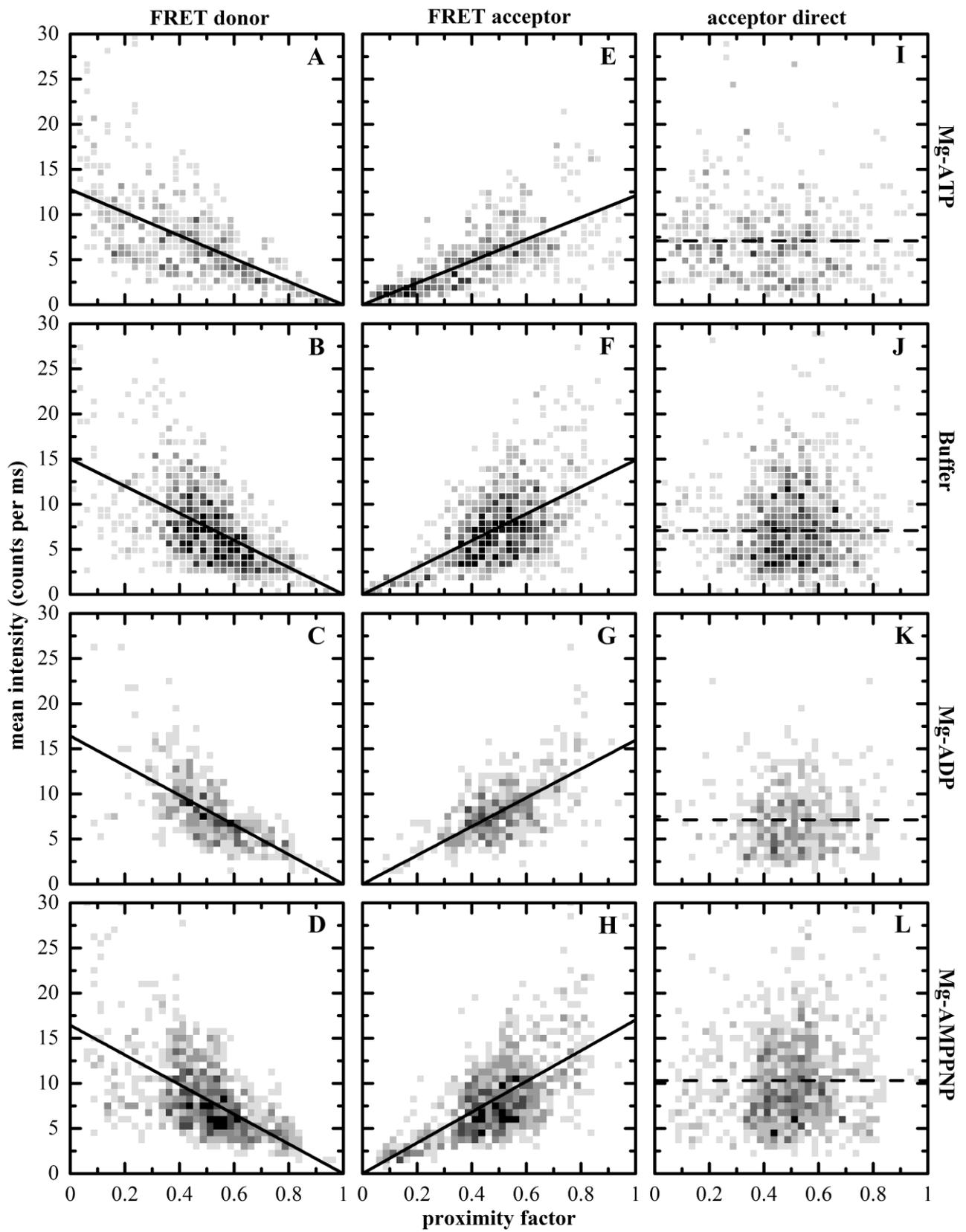

**Figure 5: A-D)** Mean intensities of FRET donor *versus* the mean proximity factor of a photon burst in the presence of 1 mM Mg-ATP, without nucleotides (buffer), 1 mM Mg-ADP, or 1 mM Mg-AMPPNP, respectively. **E-H)** Mean intensities of FRET acceptor *versus* the mean proximity factor of a photon burst in the presence of 1 mM Mg-ATP, without nucleotides (buffer), 1 mM Mg-ADP, or 1 mM Mg-AMPPNP, respectively. Black lines represent linear fittings with offsets set to 0 at P = 1 for the FRET donor or 0 at P = 0 for the FRET acceptor. **I-L)** Mean intensities of directly excited FRET acceptor versus mean proximity factor of a photon burst in the presence of 1 mM Mg-ATP, without nucleotides (buffer), 1 mM Mg-ADP, or 1 mM Mg-AMPPNP, respectively. Dashed lines represent fittings with slopes set to 0.

The mean FRET acceptor intensities following direct excitation of these FRET-labeled $A_3B_3DF$ complexes (Fig. 5I-L) were slightly higher in the presence of Mg-AMPPNP, potentially caused by a higher background. However, no deviation of the mean FRET acceptor intensities were found for the "*low FRET*" population with 0.1 < P < 0.25 in the presence of Mg-ATP. Therefore, we concluded that FRET acceptor quenching can be excluded as the cause for the "*low FRET*" population in the presence of Mg-ATP.

Fig. 6A-D shows the histograms of proximity factors for all four biochemical conditions. All four histograms show a peak with a high proximity factor between 0.3 and 0.75. Only in the presence of 1 mM Mg-ATP an additional peak appeared with a low proximity factors between 0.1 and 0.3, while at the same time the population of bursts with the high proximity factor was decreased. This is in line with the interpretation found previously that subunit F is in an extended conformation in the presence of Mg-ATP, i.e. when the enzyme is actively hydrolyzing ATP[23].

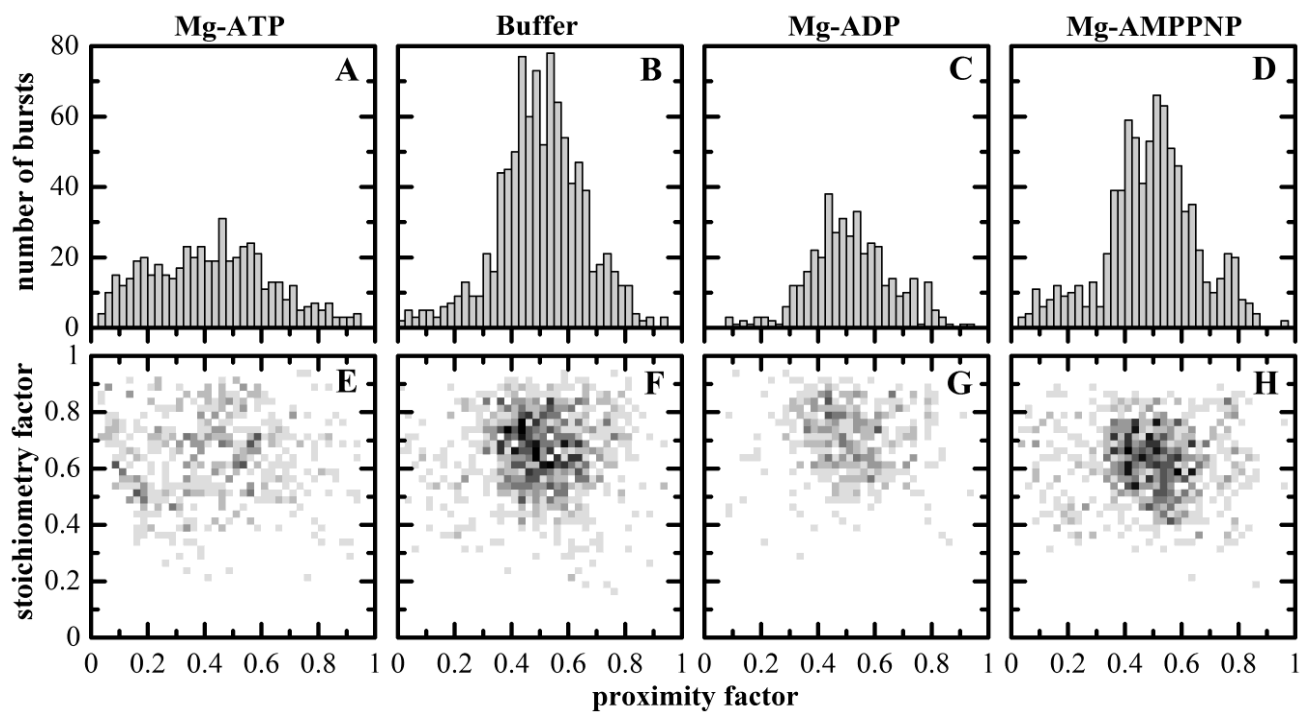

**Figure 6: A-D)** Proximity factors of FRET labeled $A_3B_3DF$ complexes in the presence of 1 mM Mg-ATP, without nucleotides (buffer), 1 mM Mg-ADP, or 1 mM Mg-AMPPNP, respectively. **E-G)** 2D-distributions of stoichiometry factor S *versus* the proximity factor P of FRET labeled $A_3B_3DF$ complexes in the presence of 1 mM Mg-ATP, without nucleotides (buffer), 1 mM Mg-ADP, or 1 mM Mg-AMPPNP, respectively.

### 3.5 Probing fluorophore quenching by the stoichiometry factor

An alternative approach to validate distance changes measured by smFRET or to identify deviations caused by fluorescence quenching is the stoichiometry factor plot[33, 34]. For our smFRET analysis control a simplified stoichiometry factor S, omitting all corrections (i.e. for spectral detection efficiencies, quantum yields, cross-talk or photon leakage of donor fluorescence into the acceptor channel, and *vice versa*) except for background-correction, was used and defined as

$$S = (I_D + I_A) / (I_D + I_A + I_{A,direct}) \qquad (2)$$

with $I_D$, FRET donor intensity by excitation with the 488 nm laser, $I_A$, FRET acceptor intensity by excitation with the 488 nm laser, and $I_{A,direct}$, FRET acceptor intensity by direct excitation with the 635 nm laser.

Fig. 6E-H shows the two-dimensional distributions of stoichiometry factor S *versus* the proximity factor P in the presence or absence of nucleotides. For all four biochemical conditions the stoichiometry factor values varied between 0.4 and 0.9. However, no evidence was found for specific quenching of the FRET acceptor fluorophore in the presence of Mg-ATP that would have caused a strong shift of stoichiometry factors to higher values for the "*low FRET*" population. Instead, the few photon bursts with low P values found in the absence of Mg-ATP (Fig. 6F-H) showed the same S values as the "*low FRET*" photon bursts in the presence of Mg-ATP (Fig. 6E).

## 4. DISCUSSION

The CTD of subunit F of $A_1A_O$-ATP synthase is enhancing ATP hydrolysis activity in the *M. mazei* Gö1 $A_3B_3DF$ complex. As the catalytic nucleotide binding side is formed at the interface of subunits A and B, the CTD of subunit F has to extend into the $A_1$ headpiece to induce this effect. It has been shown by NMR spectroscopy that the flexible CTD of soluble subunit F can exist in a retracted or extended form[22]. However, it remains unknown what causes the structural changes. We have developed a smFRET experiment to monitor the CTD movements of subunit F in the $A_3B_3DF$ complex in real time by attaching the FRET donor to subunit D and the FRET acceptor to subunit F. Analysis of the FRET efficiency histograms based on fluorescence intensities revealed an additional "*low FRET*" population in the presence of Mg-ATP, but not with Mg-ADP nor Mg-AMPPNP, or with buffer only (Fig. 6E-H). This was interpreted as an indication for an extended conformation of the CTD of subunit F in the presence of Mg-ATP[23].

Here, we aimed at confirming these smFRET results by analyzing the simultaneously recorded fluorescence lifetime data of single $A_3B_3DF$ complexes. FRET efficiency $E_{FRET}$ is either calculated from the corrected FRET donor and acceptor intensities as given in equation (1), or from the FRET donor lifetime in the presence of the FRET acceptor fluorophore, $\tau_{DA}$, and the undisturbed FRET donor lifetime in the absence of a FRET acceptor fluorophore, $\tau_D$, as follows:

$$E_{FRET} = 1 - (\tau_{DA} / \tau_D) \sim P \qquad (3)$$

Accordingly, a proximity factor $P = 0.5$ corresponds to a lifetime ratio $(\tau_{DA} / \tau_D) = 0.5$, and a proximity factor $P = 0.2$ corresponds to a lifetime ratio $(\tau_{DA} / \tau_D) = 0.8$. Given a FRET donor lifetime $\tau_D = 4.1$ ns for Atto488 in solution in the absence of any FRET acceptor fluorophore we expected to find FRET donor lifetimes for the FRET-labeled $A_3B_3DF$ complexes for the "*high FRET*" population with $0.4 < P < 0.75$ in the range of $(0.6 * 4.1$ ns$) > \tau_{DA} > (0.25 * 4.1$ ns$)$, i.e. between a maximum of 2.5 ns and a minimum of 1 ns. For the "*low FRET*" population with $P \sim 0.2$ we expected a corresponding lifetime $\tau_{DA} \sim 3.3$ ns. However, in $A_3B_3DF$ complexes exhibiting only the FRET donor fluorescence we found a double exponential decay for the combined photons from all bursts with short and long lifetime components of 0.7 ns and 4.2 ns, respectively. This was unexpected because the donor fluorescence lifetime should exhibit a single mono-exponential decay with a lifetime comparable to Atto488 in solution (i.e. 4.1 ns). The long lifetime component of the $A_3B_3DF$ complexes was in good agreement with this lifetime. In contrast, the short lifetime component represented an unknown population of the dye that was affected or quenched by the local protein environment. As a result, the FRET donor lifetimes for FRET-labeled $A_3B_3DF$ complexes classified as "*low FRET*" or as "*high FRET*", respectively, did not show lifetimes that could be fitted by monoexponential decays. Instead, double-exponential fits were required, and neither the expected donor lifetime for the "*low FRET*" nor for the "*high FRET*" population was found. For each population of similar "*low FRET*" or "*high FRET*" $A_3B_3DF$ complexes the individual single-molecule lifetimes ranged from 1.1 ns to 3.2 ns. We conclude that for our smFRET experiments the donor fluorescence lifetime cannot be used as an independent measure and independent control of the intensity-derived FRET efficiency. In contrast, for the acceptor fluorophore excited directly we observed a mono-exponential decay with a lifetime of 3.7 ns. This was in very good agreement with the published fluorescence lifetime of Atto647N in solution (3.6 ns). We conclude that the FRET acceptor fluorescence on subunit F was not disturbed by the local protein environment.

The linear relations of mean FRET donor intensity *versus* proximity factor, or of mean FRET acceptor intensity *versus* proximity factor, respectively, confirmed that the "*low FRET*" population of $A_3B_3DF$ complexes in the presence of Mg-ATP was not a result of a specific quenching of the FRET acceptor fluorophore by Mg-ATP. Additionally, the stoichiometry factor *versus* proximity factor plots did not indicate any obvious quenching of the FRET acceptor fluorophore by Mg-ATP.

The number of photons used for calculating FRET donor lifetimes and the proximity factor P in each $A_3B_3DF$ complex was limited due to Brownian motion of the freely diffusing proteins and resulted in large standard deviations for P. Mean transit times through the confocal detection volume were in the range of a few ms. In order to extend observation times and photon counts of single proteins, DNA molecules or nanoparticles in solution, A E Cohen and W E Moerner have invented a microfluidic device called the Anti-Brownian electrokinetic trap, ABELtrap[35-38]. The ABELtrap confines diffusion in two dimensions and uses a fluorescence-based localization of the single molecule within the trap region for a fast feedback to four electrodes that continuously pushes the molecule back to the center of the trap. Analyzing the applied voltages allows to extract diffusion and charge properties of the molecule[37, 39-41]. The soluble $F_1$ complex of the *E. coli* $F_1F_O$-ATP synthase was studied in the ABELtrap to monitor the regulatory conformational changes of the ε subunit by smFRET[42]. Observation times were extended by a factor of 100 in the ABEL trap compared to freely diffusing $F_1$ complexes[1,43], and proximity factors could be analyzed not only to obtain a mean P value but to identify intensity-based FRET fluctuations with a single photon burst. Holding single molecules in the ABELtrap results in large numbers of recorded photons. Using pulsed excitation allows to determine single-molecule fluorescence lifetimes with very high precision and to reconstruct lifetime fluctuations of a trapped single molecule[38,44-46]. Future measurements of conformational changes in FRET-labeled $A_3B_3DF$ complexes or in reconstituted enzymes in liposomes[47-49] will benefit from the extended observation times in an ABELtrap and by avoiding artefacts induced by surface immobilization techniques.

Summarizing the refined smFRET data analysis presented here, we are convinced that the time-resolved fluorescence lifetime measurements of the FRET-labeled $A_3B_3DF$ complexes provided valuable controls for the intensity-based observation of a conformational change of the CTD of subunit F in the presence of Mg-ATP. However, the multi-exponential FRET donor lifetime decay in the absence of a FRET acceptor (as well as in the presence) prevents a straightforward calculation of the intramolecular distance between the two FRET fluorophores in $A_3B_3DF$ complexes based on FRET theory. As a consequence we expect deviations and strong fluctuations for the mean fluorescence quantum yield of the FRET donor fluorophore bound to subunit D. Because we have not measured single-molecule spectra for both FRET fluorophores attached to the respective $A_3B_3DF$ subunits, a quantitative distance calculation is not possible. Nevertheless, the changes in FRET efficiency upon addition of Mg-ATP can be explained most likely by an increase in the relative distance, i.e. our smFRET data are strongly supporting the concept of an extended CTD conformation of subunit F in the active enzyme.

## ACKNOWLEDGEMENTS

The authors thank all members of our research groups who supported various aspects of this work. Financial support by the German Research Foundation (DFG grants BO-1891/10-2 and BO-1891/15-1 to M.B.) and by the Ministry of Health, Singapore (NMRC, CBRG12nov049 to G.G.) is gratefully acknowledged. D.S. received a research scholarship from the Ministry of Education, Singapore.